\documentclass[3p,times,twocolumn]{elsarticle}
 \biboptions{comma,sort&compress}
 
\usepackage{graphicx}
\usepackage{amsmath}
\usepackage{here}
\usepackage{ecrc}

\volume{00}

\firstpage{1}

\journalname{Nuclear and Particle Physics Proceedings}

\runauth{}

\jid{nppp}

\jnltitlelogo{Nuclear and Particle Physics Proceedings}

\usepackage{amssymb}

\usepackage[figuresright]{rotating}

\usepackage{bm}
\usepackage{accents}

\newcommand*{\dt}[1]{%
  \accentset{\mbox{\scriptsize\bfseries .}}{#1}}

\newcommand{\GeV}{\,{\rm GeV}}
\newcommand{\TeV}{\,{\rm TeV}}

\newcommand{\dil}{{\rm \mbox{{\scriptsize  dil}}}}

\begin{document}

\begin{frontmatter}

\title{Light dilatons in warped space: Higgs boson and LHCb anomalies$^*$}
 \cortext[cor0]{Talk given at 19th International Conference in Quantum Chromodynamics (QCD 16),  4 - 8 july 2016, Montpellier - FR}
 \author[label1]{Eugenio~Meg\'{\i}as\fnref{fn1}}
 \fntext[fn1]{Speaker, Corresponding author.}
 \ead{emegias@mppmu.mpg.de}
 \address[label1]{Max-Planck-Institut f\"ur Physik (Werner-Heisenberg-Institut), F\"ohringer Ring 6, D-80805, Munich, Germany}

 \author[label2]{Giuliano Panico}
 \ead{gpanico@ifae.es}

 \author[label2]{Oriol Pujol\`as}
 \ead{pujolas@ifae.es}

 \author[label2,label3]{Mariano Quir\'os}
 \ead{quiros@ifae.es}

 \address[label2]{Institut de F\'{\i}sica d'Altes Energies (IFAE), The Barcelona Institute of  Science and Technology (BIST), Campus UAB, 08193 Bellaterra (Barcelona) Spain}

 \address[label3]{ICREA, Pg. Llu\'is Companys 23, 08010 Barcelona, Spain}

\pagestyle{myheadings}
\markright{ }
\begin{abstract}
We study the extension of the Standard Model (SM) with a light dilaton in a five dimensional warped model. In particular, we analyze the coupling of the dilaton with the SM matter fields, compare the model predictions with Electroweak Precisions Tests and find the corresponding bounds on the mass of the lightest Kaluza-Klein modes. We also investigate the possibility that the Higgs-like resonance found at the LHC can be a dilaton. Finally, we show that our set-up can also provide an explanation of the anomalies recently observed in $B$-meson decays. 
\end{abstract}

\end{frontmatter}

\section{Introduction}
\label{sec:intro}

The discovery of a Higgs-like particle at the LHC in 2012, together with the theory expectation of possible new resonances not predicted by the SM~\cite{Olsen}, initiated an increasing activity in the study of Beyond the SM (BSM) physics. In this respect the Electroweak (EW) Hierarchy Problem  has become one of the most pressing theoretical issues in high energy particle physics. Several scenarios have been proposed for a BSM dynamics that could solve the Hierarchy Problem, but no unambiguous experimental evidence is yet available to discriminate among them. Indirect searches can provide some evidence for new phenomena, as in several cases they can be sensitive to scales much higher than the $\TeV$. An intriguing example are the anomalies in the $B$-meson decays recently found at the LHCb experiment~\cite{Aaij:2014ora,Aaij:2015oid}, which could be suggestive of a violation of universality in the lepton sector. Some extensions of the SM including massive vector bosons $(Z^\prime)$~\cite{Altmannshofer:2013foa,Allanach:2015gkd} have been proposed to interpret them.

In this work we try to connect the flavor anomalies with a BSM dynamics motivated by addressing the EW Hierarchy Problem. Extra-dimensional models {\it \`a la} Randran-Sundrum (RS)~\cite{Randall:1999vf} provide a natural way to achieve this aim, as they predict new massive Kaluza-Klein (KK) vector modes that can give rise to new effective interactions among the SM fermions. These models also allow the possibility that the SM is part of a nearly-conformal sector~\cite{Contino,Bellazzini:2013fga,Megias:2014iwa,Megias:2015qqh}. This is a very intriguing scenario as well, since it leads to the appearance of a light dilaton that could be either identified with the Higgs or provide a new resonance which affects the Higgs-sector phenomenology.

\section{The model}
\label{sec:model}

We introduce a scenario that is analogous to the usual RS set-up. We consider an extra-dimensional model in which the extra-dimension is close to AdS$_5$ near the UV brane, the only difference with respect to the RS set-up being a deformation of the background metric near the IR boundary. The  metric in 5 dimensions (5D) is~$ds^2 = e^{-2A(y)} \eta_{\mu\nu} dx^\mu dx^\nu + dy^2$, where $\eta_{\mu\nu}= (-1,1,1,1)$, $y$ is the proper coordinate along the extra-dimension, and the action of the model is
\begin{eqnarray}
&&\hspace{-1cm} S_\phi =  M^3\int d^4xdy\sqrt{-g}\left(R-\frac{1}{2}(\partial_M \phi)^2-V(\phi)\right) \nonumber \\
&&\hspace{-1cm}\qquad -M^3 \sum_{\alpha}\int d^4x dy \sqrt{-g}\,2\mathcal V^\alpha(\phi)\delta(y-y_\alpha) \,,  \label{eq:model}
\end{eqnarray}
where $\phi$ is the scalar field, ${\mathcal V}^\alpha \; (\alpha=0,1)$ are the UV and IR 4 dimensional (4D) brane potentials at $y_0 \equiv y(\phi_0)$ and $y_1 \equiv y(\phi_1)$ respectively, and $M$ is the 5D Planck scale.

The second-order equations of motion of the model can be reduced to first-order form by introducing a superpotential $W(\phi)$~\cite{Gubser:2000nd}, given by~$V(\phi) \equiv \frac{1}{2} [W^\prime(\phi)]^2 - \frac{1}{3} W(\phi)^2$. The background equations of motion then reduce to~$\dot{A}(y) = \frac{1}{6} W(\phi(y))$ and $\dot{\phi}(y) =  W^\prime(\phi)$, where we use the notation $\dot{X}\equiv dX(y)/dy$, and $Y^\prime\equiv dY(\phi)/d\phi$. 
The localization of the branes turns out to be governed by the effective potentials~$U_\alpha(\phi)\equiv\mathcal V_\alpha(\phi)-(-1)^\alpha W(\phi)$. The boundary conditions together with the equations of motion lead to $U_\alpha(\phi)\big|_{y=y_\alpha} =U_\alpha^\prime(\phi)\big|_{y=y_\alpha}=0$. In order to solve the Hierarchy Problem, the brane dynamics should fix $(\phi_0,\phi_1)$ to get $A(\phi_1) - A(\phi_0) \approx 35$, as this implies $M_{\textrm{\scriptsize Planck}} \simeq 10^{15} M_{\textrm{\scriptsize TeV}}$. We will consider $\phi_1=5$, while $\phi_0$ is used to fix the length of the extra-dimension. In the following we assume the dynamics of $\phi$ to be characterized by the analytic superpotential
\begin{equation}
W(\phi)=6k \left( 1+e^{a\phi} \right)^b \,, \label{eq:W}
\end{equation}
where $a$ and $b$ are dimensionless parameters controlling the background, and $k$ is a mass dimension parameter related to the curvature along the fifth dimension~\cite{Cabrer:2009we}.

\section{Electroweak breaking}
\label{sec:electroweak}

In order to introduce the electroweak sector in the theory, we assume a 5D gauge invariance, whose gauge group concides with the SM one $\textrm{SU}(2)_L \times \textrm{U}(1)_Y \times \textrm{SU}(3)_c$. We denote the corresponding gauge fields as $W_M^a(x,y)$, $B_M(x,y)$, $G_M^A(x,y)$, where $M=(\mu,5)$, $a=1,2,3$ and $A=1,\dots,8$. In addition, we consider a Higgs field propagating in the bulk, and we write it as
\begin{equation}
H(x,y)=\frac{1}{\sqrt{2}}e^{i\chi(x,y)}\left( \begin{array}{c}  0\\ h(y)+\xi(x,y) \end{array}   \right)\,,
\end{equation}
where $h(y)$ is the Higgs background, $\xi(x,y)$ are physical fluctuations, and $\chi(x,y)$ are Goldstone modes. The action of the model is then $S = S_\phi + S_5$ with
\begin{eqnarray}
&&\hspace{-1.5cm} S_5 = \!\! \int \!\! d^4x dy\sqrt{-g}\left(-\frac{1}{4} \vec W^{2}_{MN}-\frac{1}{4}B_{MN}^2-|D_M H|^2-V(H)\right) \,, \nonumber  \\
  && \label{eq:S5}
\end{eqnarray}
where $V(H)$ is the 5D Higgs potential. EW symmetry breaking (EWSB) is triggered by an IR brane potential. The Higgs bulk potential contains a mass term $M^2(\phi)=\alpha k \left[\alpha k-\frac{2}{3}W(\phi)  \right]$. The parameter $\alpha$ controls the localization of the Higgs and thus is connected to the amount of tuning related to the Hierarchy Problem~\cite{Cabrer:2011fb}.

\begin{figure*}[htb]
\centering
  \begin{tabular}{c@{\hspace{2.5em}}c}
    \includegraphics[width=.445\textwidth]{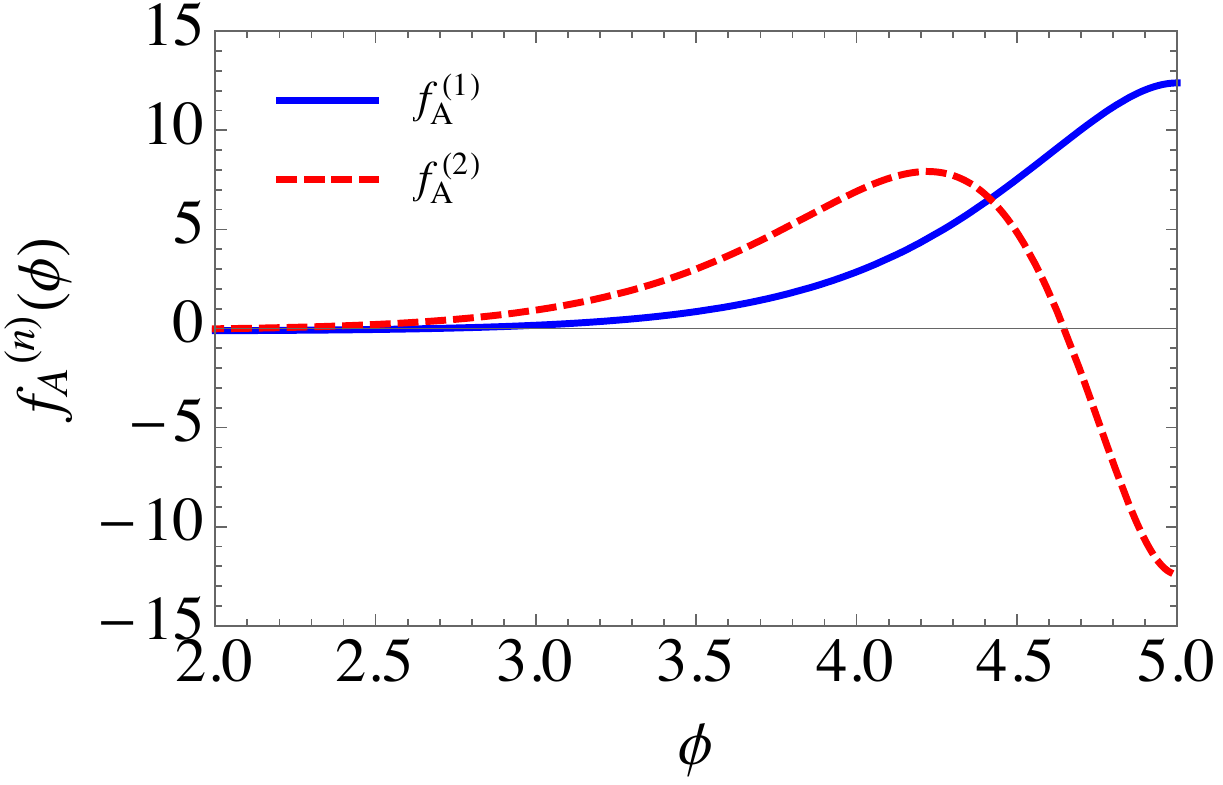} &
     \includegraphics[width=.435\textwidth]{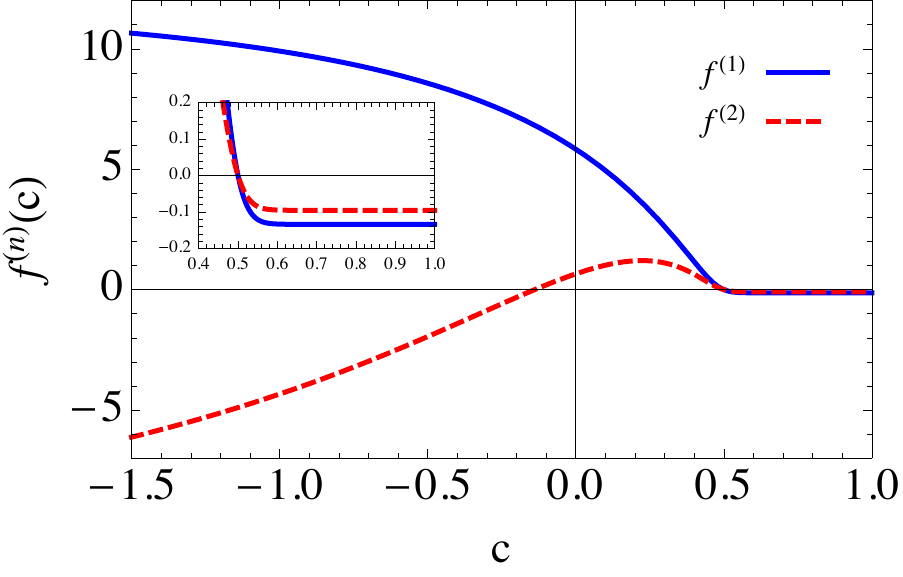}
\end{tabular}
\vspace{-0.4cm}
\caption{\it (Left) Profiles of the gauge boson KK modes $f_A^{(n)}$ for $n=1,2$ (solid blue and dashed red lines respectively). (Right) Coupling (normalized with respect to the 4D coupling $g$) of a fermion zero-mode with the $n$-th KK gauge field, $f^{(n)}(c)$, as a function of the fermion localization parameter $c$ (cf.~Eq.~(\ref{eq:fn})). In both panels we have considered $b=1$, $a=0.2$ in Eq.~(\ref{eq:W}).}
\label{fig:KKcouplings}
\end{figure*}

The gauge fields can be decomposed in KK modes as $A_\mu(x,y)= \sum_n f^{(n)}_A(y) A^{n}_\mu(x)/\sqrt{y_1}$, where $f_A^{(n)}(y)$ satisfies Neumann boundary conditions and bulk equations~$\left(m^{(n)}_A\right)^2f^{(n)}_A+\left( e^{-2A}\dot{f}^{(n)}_A \right)^{\dt{}}-M_A^2(y) f^{(n)}_A=0$, where $m_A^{(n)}$ denotes the mass of the $n$-th KK mode and $M_A(y)$ is the mass term induced by the vacuum expectation value of the Higgs after EWSB~\cite{Cabrer:2011fb}.  We plot $f_A^{(n)}$ in Fig.~\ref{fig:KKcouplings} (left).

\begin{figure}[htb]
\centering
      \includegraphics[width=.40\textwidth]{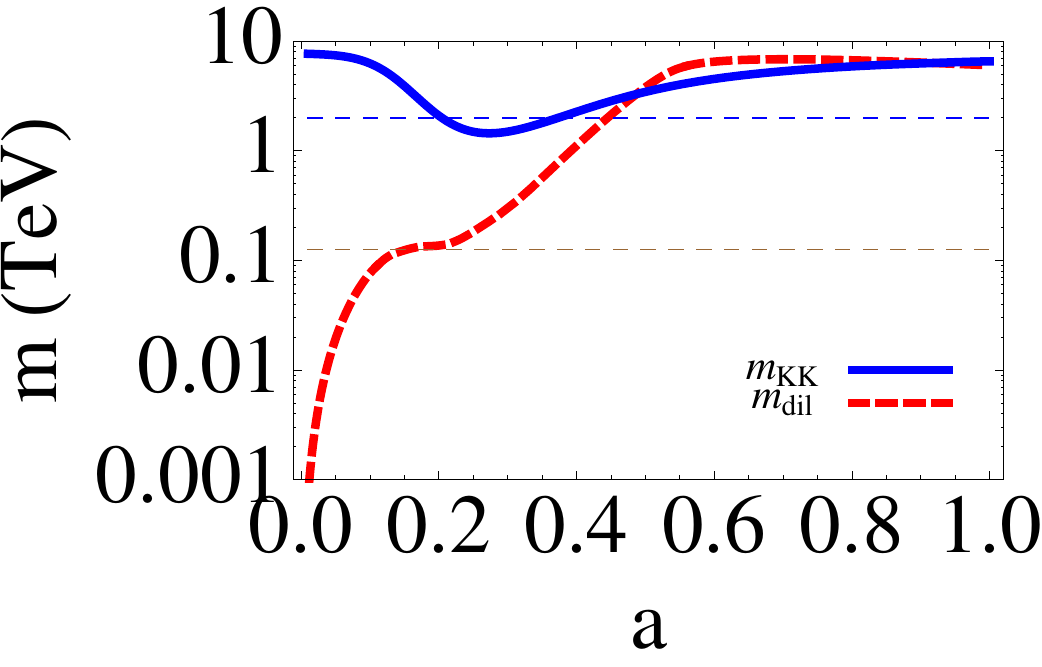}
\vspace{-0.4cm}
\caption{\it Lower bound on KK mass (solid line) as a function of the parameter $a$, computed from electroweak observables. The corresponding dilaton mass is in dashed line. We have drawn horizontal dashed lines corresponding to $125\GeV$ and $2\TeV$.}
\label{fig:bound}
\end{figure}

The SM fermions are realized as chiral zero modes of 5D fermions. The localization of the different fermions is determined by the 5D mass terms $M_{f_{L,R}}(y)=\mp c_{f_{L,R}} W(\phi)$~\cite{Cabrer:2011qb}. The zero modes are localized near the UV (IR) brane for $c_{f_{L,R}}>1/2$ ($c_{f_{L,R}}<1/2$). A value $c_{f_{L,R}} < 1/2$ thus corresponds to a sizable amount of compositeness for the corresponding fermions, whereas $c_{f_{L,R}} > 1/2$ characterizes fermions that are almost elementary. An important ingredient in Sec.~\ref{sec:LHCbanomalies} is the coupling of the SM fermions with the massive KK modes of the gauge fields, which are universal and fully determined by the localization of the fermions, i.e. by the $c_{f_{L,R}}$ parameters. The coupling with the $n$-th gauge KK mode, $X_\mu^n$, can be written as~$g_{f_{L,R}}^{X^n}\,X_\mu^n \bar f_{L,R}\gamma^\mu f_{L,R} \equiv g f^{(n)}(c_{f_{L,R}})X_\mu^n\, \bar f_{L,R}\gamma^\mu f_{L,R}$, where $f_{L,R}$ are fermion zero-modes, $g$ is the SM gauge coupling and
\begin{equation}
\hspace{-0.3cm} f^{(n)}(c)=\frac{\sqrt{y_1}}{\sqrt{\displaystyle \int_0^{y_1} \!\! \left(f_A^{(n)}(y)\right)^2}}\frac{\displaystyle \int_0^{y_1} \!\! f_A^{(n)}(y) e^{(1-2c)A(y)}}{\displaystyle \int_0^{y_1}\! e^{(1-2c)A(y)}}\,. \label{eq:fn}
\end{equation}
These functions are plotted in Fig.~\ref{fig:KKcouplings} (right). Note that for fields which are almost elementary the coupling becomes rather weak $\sim 0.1 g$. When comparing the model predictions with EW precision tests, the most relevant bounds come from the universal oblique observables, encoded by the $(S,T,U)$ variables defined in Ref.~\cite{Peskin:1991sw}. These constraints imply a lower bound on the mass of the vector KK modes as well as on the mass of the scalar mode (the dilaton) for a fixed potential $U_\alpha(\phi)$. The results are showed in Fig.~\ref{fig:bound}. We find that for $a \sim 0.25$ the KK-modes are allowed to have a mass $m_{KK} = {\cal O}(\TeV)$ and the dilaton is quite light, $m_\dil \lesssim \mathcal O(100)\GeV$. 

\section{Coupling of dilaton to SM matter fields}
\label{sec:couplings}

In view of the results of the previous section, a natural question arises: Is the $125\GeV$ Higgs resonance found at LHC actually consistent with a dilaton? In order to answer this, it is not enough to study the mass of the dilaton, but also the couplings of dilaton with SM matter fields. A comparison with the couplings predicted for the Higgs in the SM would shed light on this problem.

If we consider a scalar perturbation in the metric $ds^2 \sim e^{-2(A+F)}$, and we decompose the scalar fluctuation as $F(x,y) = F(y) {\cal R}(x)$, we can compute the coupling of the dilaton to gauge bosons and fermions~\cite{Megias:2015ory}:
\begin{eqnarray}
  &&\hspace{-1.5cm} \mathcal L_{rad}=\frac{r(x)}{v}\bigg\{ \frac{\alpha_{EM}}{8\pi}(b_1+b_{1/2})\,c_\gamma\,F_{\mu\nu}F^{\mu\nu}-2 c_W\,m_W^2 W_\mu W^\mu \nonumber \\
  &&-c_Z\,m_Z^2 Z_\mu Z^\mu-c_f m_f \bar f f \bigg\} \,, \label{eq:LV}
\end{eqnarray}
where $r(x)$ is the canonically normalized radion field. The values $c_\gamma = c_W = c_Z = c_f = 1$ correspond to the SM Higgs couplings. After an expansion of the action of the model to linear order in perturbations, one gets a value of the couplings of order ${\cal O}(10^{-4}) -  {\cal O}(10^{-2})$. The couplings are very small, so that the dilaton can not play the role of a Higgs impostor. Its presence, however, only leads to small deviations in SM predictions of the same tiny order~\cite{Megias:2015ory,Megias:2015qqh}. An extension of this model, including higher dimensional operators that could account for ``strongly coupled" dilatons ($c_X\gtrsim 1$, $X=\gamma,W,Z,f$) has been studied in~\cite{Megias:2015ory}. However, even in that case, the possibility of a Higgs impostor has been disfavored. In conclusion, we find two scenarios: {\it i) dilaton extension of the SM:} $c_{X} \ll 1$ and $m_\dil < 100\GeV$, in which the dilaton would be the first new state in the spectrum in addition to all the SM particles (including the Higgs), and {\it ii) a Higgs impostor:} $c_{X} \sim 1$ and $m_\dil \sim 125\GeV$, in which the dilaton would be the already discovered $125\GeV$ resonance. The first scenario is rather natural while, in the class of models studied here, the second scenario seems to be strongly disfavoured.

\section{The LHCb anomalies}
\label{sec:LHCbanomalies}

Recent results found by the LHCb collaboration in $B$-meson decays seem to point toward the existence of new physics. The LHCb measurements of the angular distribution in the decay $B\to K^*\mu^+\mu^-$ and the ratio of branching fractions ${\rm BR}(B \rightarrow K \mu^+ \mu^-)/{\rm BR}(B \rightarrow K e^+ e^-) \simeq 0.745^{+0.090}_{-0.074}\pm 0.036$~\cite{Aaij:2014ora,Aaij:2015oid}, which differs from the SM prediction at the $2.6\sigma$, could be suggestive of a violation of universality in the lepton sector. We will check if our model can accommodate this anomaly.

\subsection{The $B\to K^*\mu^+\mu^-$ anomaly}
\label{sec:Banomaly}

\begin{figure}[htb]
\centering
 \includegraphics[width=.35\textwidth]{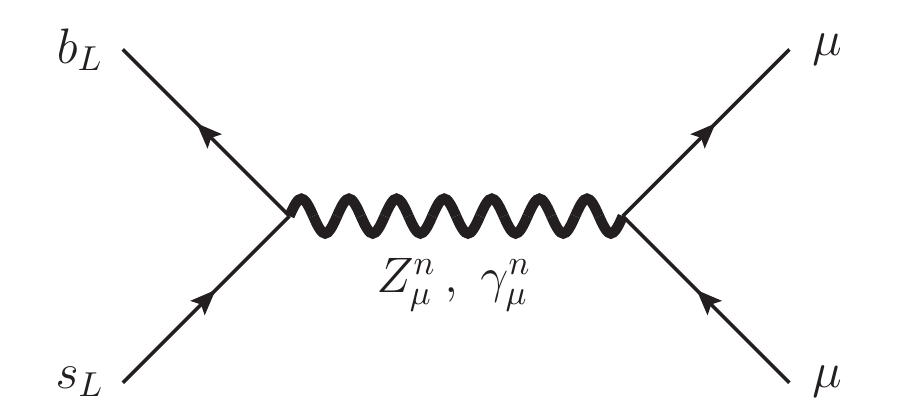}
\vspace{-0.4cm}  
\caption{\it Diagrams giving rise to the effective operators of Eq.~(\ref{operators}).}
\label{fig:c9}
\end{figure}

After EWSB the relevant four-fermion effective operators contributing to $\Delta F = 1$ transitions are~\cite{Buchalla:1995vs}
\begin{equation}
\mathcal L_{eff}=\frac{G_F \alpha_{EM}}{\sqrt{2}\,\pi} V^*_{ts}V_{tb}\sum_i C_i\,\mathcal O_i\,,
\label{efflagrangian}
\end{equation}
with Wilson coefficients $C_i=C_i^{SM}+\Delta C_i$, and operators
\begin{equation}
\hspace{-0.5cm} \mathcal O_9 =(\bar s_L\gamma_{\mu}  b_L)(\bar\mu \gamma^\mu\mu)\,, \quad \mathcal O_{10}=(\bar s_L\gamma_{\mu}  b_L)(\bar\mu \gamma^\mu\gamma_5\mu)\,.
\label{operators}
\end{equation}
In the model of Secs.~\ref{sec:model} and~\ref{sec:electroweak} contact interactions involving SM fermions can be generated by the exchange of the KK modes of the $Z$-boson and of the photon, see Fig.~\ref{fig:c9}.
In this scenario, lepton universality can be broken provided that the localization of the various lepton generations is different. Under reasonable assumptions, the leading flavor violating interactions with the vector KK modes have the form~\cite{Megias:2016bde}
\begin{eqnarray}
&&\hspace{-1.0cm} \mathcal L_{EW}=\sum_{X=Z,\gamma}\, \frac{X^n_\mu}{2c_W}\Big[
    V_{3i}^* V_{3j}\,\bar d_i\gamma^\mu \bigg\{\left(g^{X^n}_{b_{L}}-\overline g^{X^n}_{L}  \right)P_L \nonumber \\
&&\hspace{1.5cm}+\left(g^{X^n}_{b_R}-\overline g^{X^n}_{R}  \right)P_R\bigg\} d_j + {\rm h.c.} \Big]\,,\label{eq:lagrangianEW_od}
\end{eqnarray}
where $c_W \equiv \cos \theta_W$, $P_{RL} = (1 \pm \gamma_5)/2$, $V_{ij}$ are the CKM matrix elements, $d_i$ ($i=1,2,3$) denotes the down-type quark in the $i$-th generation, and $\overline g^{X^n}_{L}$ are the couplings of $d_1$ and $d_2$ to the KK vectors. The couplings in Eq.~(\ref{eq:lagrangianEW_od}) give rise to the contribution to the $C_9$ Wilson coefficient
\begin{equation}
\Delta C_9=-\sum_{X=Z,\gamma}\,\sum_n\frac{\pi g_{\mu_V}^{X_n}\left(g^{X_n}_{b_L}-g^{X_n}_{s_L}\right)}{2\sqrt{2} G_F \alpha_{EM} c_W^2 M^2_n}\,.
\end{equation}
The expression for $\Delta C_{10}$ is analogous and can be obtained with the change $g_{\mu_V}^{X_n} \to g_{\mu_A}^{X_n}$. The largest contributions come from the exchange of the first KK excitations, $Z_\mu^1$ and $\gamma_\mu^1$. The additional contributions are suppressed by the larger masses of the higher states, and lead to subleading corrections.

\subsection{Constraints: $Z\overline b b$ and $Z \overline \mu \mu $ couplings}
\label{sec:ZbbZmumu}

\begin{figure}[htb]
\centering
 \includegraphics[width=0.47\textwidth]{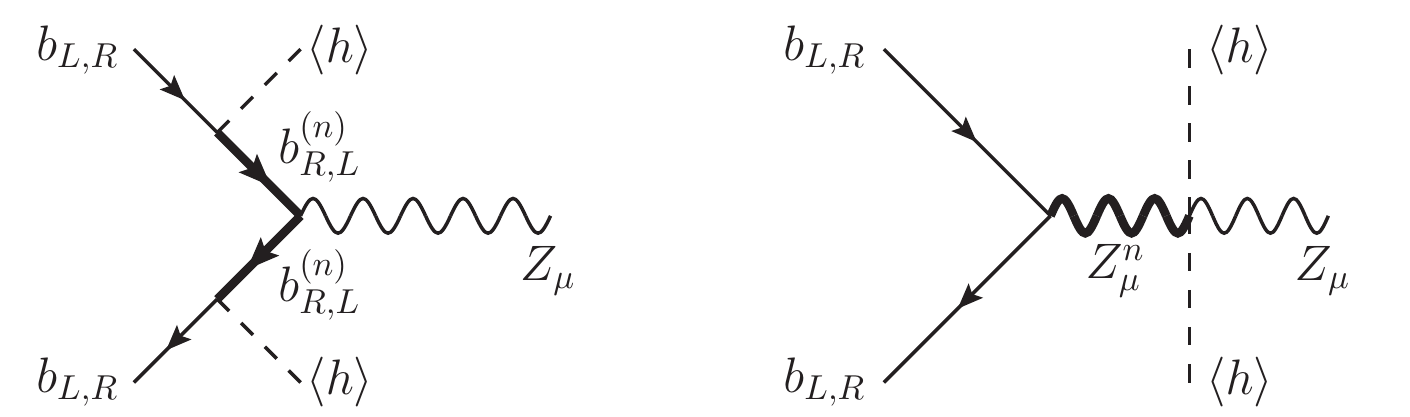} 
\vspace{-0.6cm} 
\caption{\it Diagrams contributing to  the coupling $Z\bar b b$. }
\label{fig:Zbbdiagram}
\end{figure}

\begin{figure}[htb]
\centering
 \includegraphics[width=0.38\textwidth]{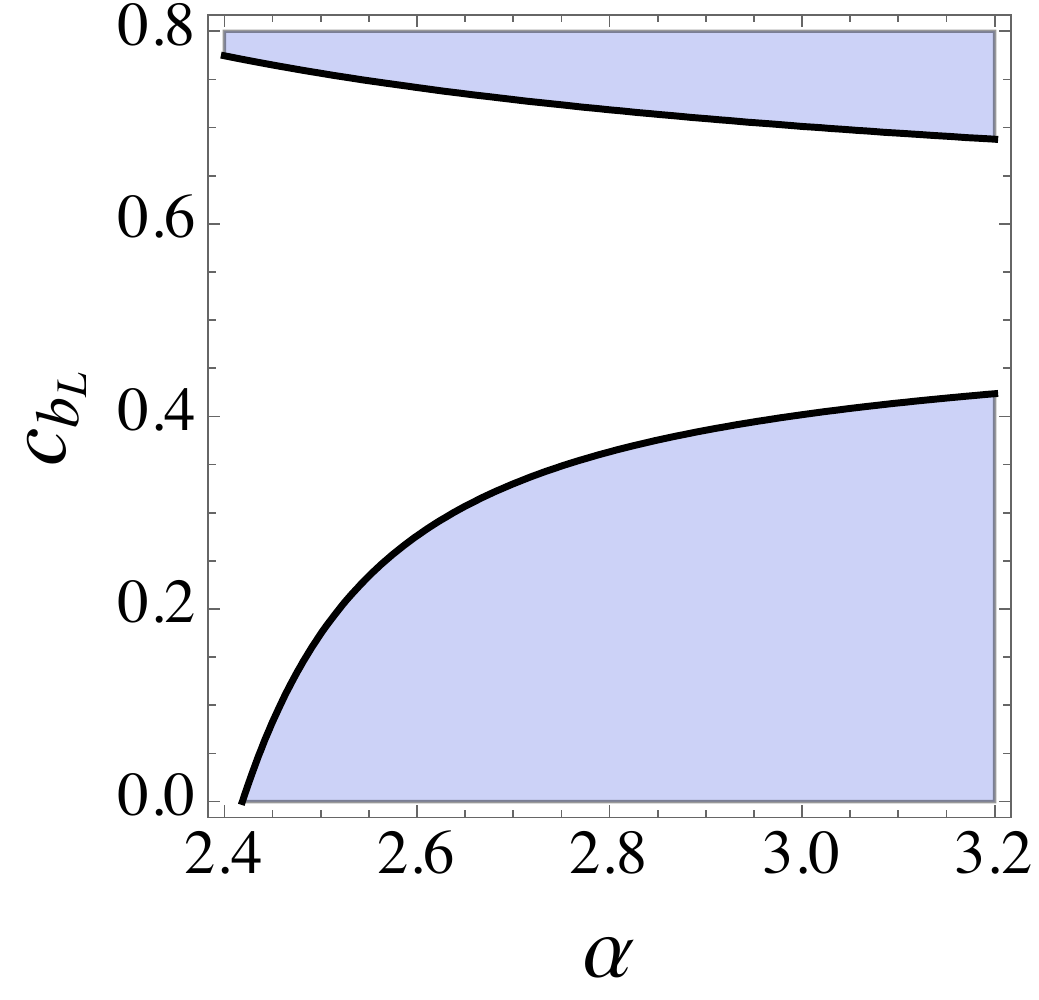} 
\vspace{-0.4cm}
 \caption{\it Region in the plane $(\alpha,c_{b_L})$ allowed by experimental data on $\delta R_b^{exp} = 0.00053\pm 0.00066$ at the $3\sigma$ level. We have fixed $c_{b_R}=0.58$.
  }
\label{fig:dRbdAb1}
\end{figure}

The $Z$ boson couplings to SM fermions can be modified by vector KK modes and fermion KK excitations. In the case of the $Z_\mu \bar b_{L,R}\gamma^\mu b_{L,R}$ couplings, the diagrams induced by these effects are shown in Fig.~\ref{fig:Zbbdiagram}.
After summing over the KK levels, the full result reads
\begin{equation}
\delta g_{b_{L,R}}= - g_{b_{L,R}}^{SM}m_Z^2\widehat \alpha_{b_{L,R}}\pm g\frac{v^2}{2}\widehat\beta_{b_{L,R}}\,,  \label{eq:delta_g}
\end{equation}
where $\widehat\alpha_{b_{L,R}}$ and $\widehat\beta_{b_{L,R}}$ are defined in~\cite{Cabrer:2011qb}. The experimental constraints on the $Z\overline b b$ coupling come from observables like $R_b$, defined as the ratio of the $Z\to \bar b b$ partial width to the inclusive hadronic width, and $A_{FB}^b$, the forward-backward asymmetry of the bottom quark~\cite{Agashe:2014kda}. We show in Fig.~\ref{fig:dRbdAb1} how the bounds on $c_{b_L}$ vary as a function of the parameter $\alpha$, which determines the amount of tuning in the Higgs sector. Values $\alpha \gtrsim 3$ correspond to a completely Natural theory, while $\alpha < 3$ correspond to exponentially large tuning. Analogously to $Z\overline b b$, the massive KK modes also induce modification on the muon couplings. The result is obtained from Eq.~(\ref{eq:delta_g}) with obvious substitutions. The current bounds on the distortions of the muon couplings to the $Z$ are given in~\cite{Agashe:2014kda}. This leads to a bound on the $\mu_L$ coupling given by $\left|\delta g_{\mu_{L}}(c_{\mu_{L}})/g^{SM}_{\mu_{L}}\right|\lesssim 5\times 10^{-3}$ at $95\%$~CL. Requiring our model to be completely natural implies~$c_{\mu_L}\gtrsim 0.4$.

\subsection{Flavor observables}
\label{sec:flavor}

\begin{figure}[htb]
\centering
  \includegraphics[width=0.38\textwidth]{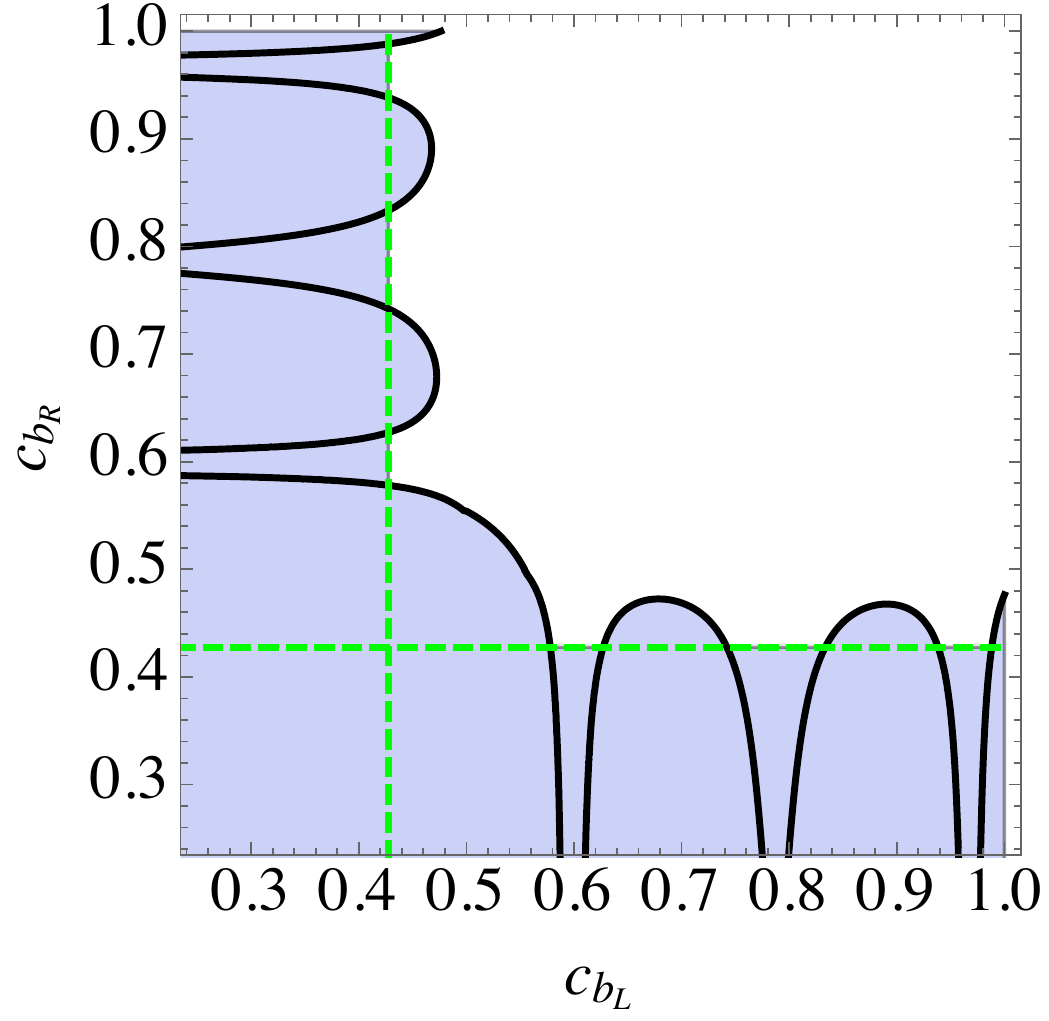}
  \vspace{-0.4cm}
\caption{\it Region in the plane $(c_{b_L},c_{b_R})$ that accommodates the bounds of Eq.~(\ref{eq:boundBs}). The dashed green lines
represent $c_{b_{L,R}} = 0.43$. The allowed points correspond to the unshaded region.
}\label{fig:flavor}
\end{figure}

Another important set of constraints comes from $\Delta F = 2$ flavor-changing processes mediated by four-fermion contact interactions. The main new physics contributions to these processes come from the exchange of gluon KK modes. After computation of the flavor-violating couplings of the KK gluons, the current bounds on the $\Delta F = 2$ contact operators~\cite{Isidori:2015oea} can be translated into constraints on the quantities
\begin{equation}
\hspace{-0.6cm} \sum_n \frac{\left(g_{b_{L,R}}^{G^n}\right)^2}{M_n^2[{\rm TeV}]}  \leq 0.14 \,, \quad \sum_n \frac{g_{b_L}^{G^n} g_{b_R}^{G^n}}{M_n^2[{\rm TeV}]} \leq 3\times10^{-4} \,. \label{eq:boundBs}
\end{equation}
The first constraint leads to $c_{b_{L,R}} \ge 0.43$. The allowed configurations in the $(c_{b_L}, c_{b_R})$ plane are shown in Fig.~\ref{fig:flavor}.

\subsection{Results}
\label{sec:results}

\begin{figure}[!htb]
\centering
    \includegraphics[width=0.38\textwidth]{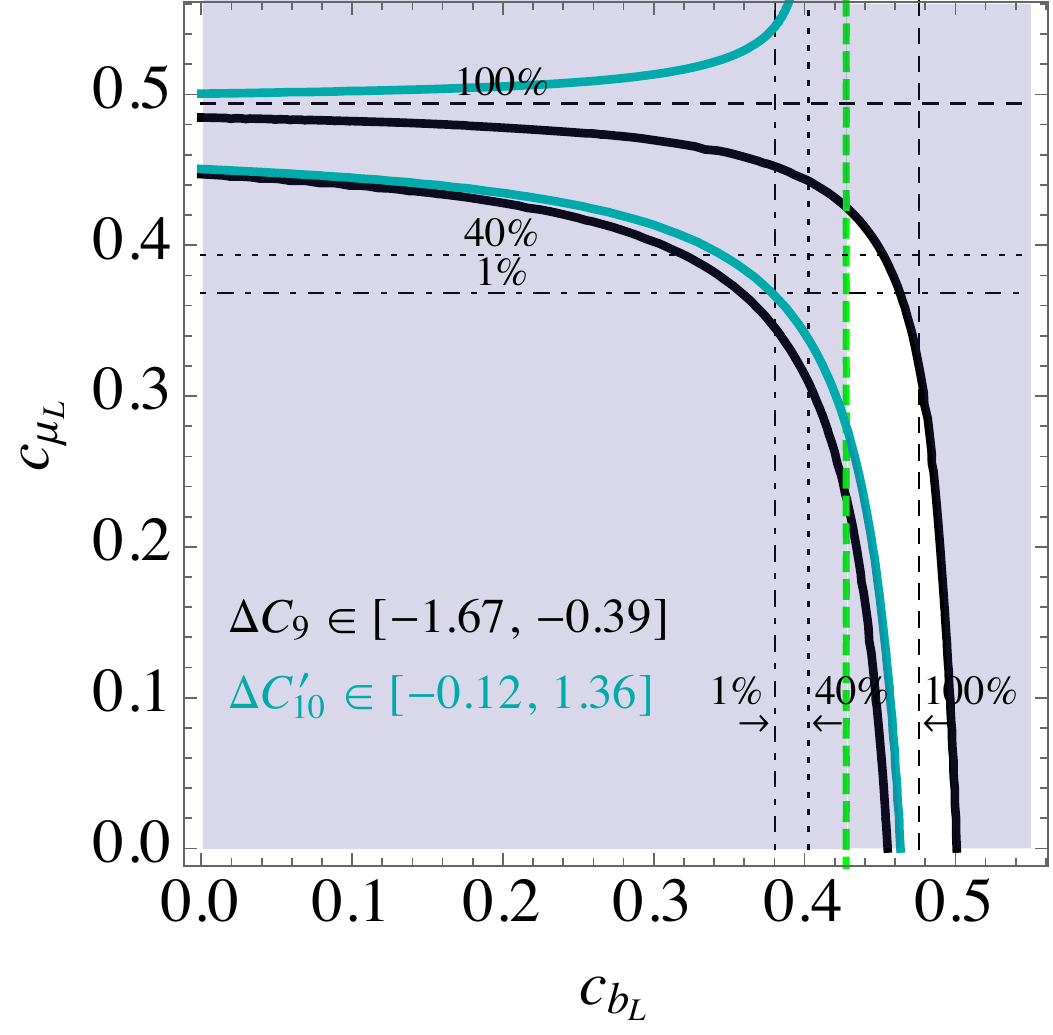} 
\vspace{-0.4cm}
\caption{\it Region in the plane $(c_{b_L},c_{\mu_L})$ that accommodates $\Delta C_9$ (the band inside the black solid lines) and $\Delta C_{10}$ (the band inside the green solid lines). The region to the left of the vertical dashed green line is excluded by Eq.~(\ref{eq:boundBs}). The fine-tuning needed to pass the constraints on the modification of the $Z\mu\bar \mu$ ($Z\bar bb$) coupling
is shown by the black dashed, dotted and dot-dashed horizontal (vertical) lines.
}
\label{fig:LHCbfinal}
\end{figure}

The results of our analysis in this section are summarized in Fig.~\ref{fig:LHCbfinal}. We show the parameter space that allows to fit the flavor anomalies, and the other constraints studied above. The horizontal and vertical black lines show the amount of fine tuning in the Higgs sector to pass the EW constraints. A completely natural scenario corresponds to $100\%$, whereas lines of $40\%$ and $1\%$ lead to a certain level of tuning.

\section{Conclusions}
\label{sec:conclusions}

We have studied an extension of the SM based on a modified RS model in 5D which leads to the existence of a dilaton whose mass can naturally be $\mathcal O(100)\GeV$. A study of the coupling to gauge fields suggests that this scenario could naturally account for new states in the spectrum lighter than the Higgs and ``weakly coupled" to the SM fields. After inclusion of reasonable higher dimensional operators, the model could potentially account for a heavier dilaton ``strongly coupled" to the SM fields. The model can as well explain the LHCb anomalies found in semi-leptonic $B$-meson decays. The flavor anomalies can be easily reproduced by assuming that the bottom and muon fields have a sizeable amount of compositeness, while the electron is almost elementary. We have found a correlation between flavor anomalies and corrections in the EW observables as well as in flavor-changing processes, that can be potentially testable in near-future experiments.

\section*{Acknowledgments} 
Work supported by MINECO Grant CICYT-FEDER-FPA2014-55613-P, Severo Ochoa Excellence Program Grant SO-2012-0234, and by Generalitat de Catalunya Grant 2014 SGR 1450. The research of E.M. is supported by the European Union (FP7-PEOPLE-2013-IEF) project PIEF-GA-2013-623006.


\end{document}